\documentstyle[aps]{revtex}
\input{epsf.sty}\input{psfig.sty}
\begin{document}
\title{Breakup of the Hydrogen Bond Network In Water:The Momentum Distribution of the Protons}
\author{G.F.Reiter}
\address{Physics Department, University of Houston, Houston,  Texas, USA}
\author{J. C. Li}
\address{Department of Physics, University of Manchester, PO Box 88,M60 1QD ,UK}
\author{J.  Mayers and T. Abdul-Redah}
 \address{ISIS Facility,  Rutherford
-Appleton Laboratory, Chilton, Didcot, OX11 0QX, UK}
\author{P.  Platzman}\address{Bell Labs Lucent Technologies, Murray Hill, New Jersey,
USA}

 \maketitle
 \begin{abstract}
  Neutron Compton Scattering
measurements presented here of the momentum distribution of
hydrogen in water at temperatures slightly below freezing to the
supercritical phase show a dramatic change in the distribution as
the hydrogen bond network becomes more disordered. Within a single
particle interpretation, the proton moves from an essentially
harmonic well in ice to a slightly anharmonic well in room
temperature water, to a deeply anharmonic potential in the
supercritical phase that is best described by a double well
potential with a separation of the wells along the bond axis of
about .3 Angstroms. Confining the supercritical water in the
interstices of a C60 powder enhances this anharmonicity. The
changes in the distribution are consistent with gas phase
formation at the hydrophobic boundaries.
\end{abstract}
\vskip 12pts PACS numbers, 61.12.Ex, 61.25.Em,82.30.Rs \vskip .5in

Although water has been intensively studied with neutron
scattering for decades, and much is known about the spatial
structure of the various phases of water, Deep Inelastic Neutron
Scattering measurements,(or Neutron Compton Scattering) which
measure the momentum distribution of the proton, have only become
available in the last few years. This distribution, which is
determined almost entirely by quantum effects for the systems we
have studied, is a sensitive probe of the local environment of the
proton, as well as a direct measurement of its dynamics.  In
crystals, for instance, measurements of the shape of the momentum
distribution  have been used to extract the Born-Oppenheimer
potential for the proton, and its changes with
temperature.\cite{rmp}  Here we demonstrate that detailed
information can be obtained from powder samples and liquids as
well. In particular, we examine the momentum distribution of the
protons in water as the regular hydrogen bond network in
poly-crystalline ice is disordered, first by melting, then by
heating under pressure into the supercritical phase, and then
further by embedding the supercritical water in the
interstices(typical diameter 100 Angstroms) of a C60 powder. We
find that there are in fact dramatic changes, as the momentum
distribution responds to changes in the  structure of the hydrogen
bonded network. The changes upon melting seem well understood, but
the results in the supercritical phase are astonishing, as the
proton appears to be coherently distributed over two positions
along the hydrogen bond separated by about .3 angstroms. The
effect of the C60 is consistent with other experiments that show a
depleted phase forming at hydrophobic surfaces.

  The experiments are done on the electron volt
spectrometer, Vesuvio, at ISIS, the pulsed neutron source at the
Rutherford Laboratory. This sort of source is needed to provide
high energy neutrons(5-100 ev) for which the energy transfer is
sufficiently large compared to the characteristic energies of the
system that the scattering is given accurately by the
impulse\cite{pmp} approximation limit.  The scattering at these
energies is entirely incoherent, each particle scattering
independently. The scattering cross-section is proportional to
$S_M(\vec{q},\omega$), the scattering function for a particle of
mass M, which is related to the momentum distribution of the
particle n($\vec{p}$) in this limit by the relation

\begin{equation}
S_M(\vec{q},\omega) = \int n(\vec{p})\delta(\omega-{\hbar
q^2\over2M}-{\vec{p}.\vec q\over M}) d\vec{p} \label{sqw}
\end{equation}
where $\hbar\omega$ is the energy transfer, M is the mass of the
proton, and q=$|\vec{q}|$ is the magnitude of the  wave-vector
transfer.  The probability of observing the proton with momentum
$\vec{p}$,$n(\vec{p})$, for simple one particle systems  in their
ground state, is the square of the absolute value of the Fourier
transform of the spatial wave-function. Although there are
certainly many body effects in water, the small mass of the proton
compared to the surrounding oxygen make this a good first
approximation for interpreting the data.\cite{wls}

 We represent
$S_M(\vec{q},\omega)$ as ${M\over q} J(\hat {q}, y)$ where
y=${M\over q}(\omega-{\hbar q^2\over 2M})$. The general case,
where the momentum distribution results from the proton in a
crystal environment, and hence has a 3-D structure, is described
in \cite{rmn}. In the present situation, where the sample is
either poly-crystalline or a liquid, the average momentum
distribution has no angular dependence, and $J(\hat {q}, y)$ is
independent of $\hat {q}$.

We will fit the data with a series expansion of the form\cite{as}
\begin{equation}
J( y) = {e^{{-y^2}\over {2\sigma^2}}\over {\sqrt{2\pi}\sigma}}
\sum\limits_{n} {a_{n}\over{2^{2n}n!}}
H_{2n}({y\over{\sqrt2\sigma}})
 \label{exp}
 \end{equation}
 where the $H_n(y)$ are Hermite polynomials.

  This series is truncated at some order(2n=14) in this case).
  The coefficients $a_{n}$  then  determine the
 measured $n(p)$ directly as a series in Laguerre polynomials.
\begin{equation}
 \label{inv}
 n(p) = {e^-{{p^2\over{2\sigma^2}}}\over{(\sqrt{2\pi}\sigma)^3}}
 \sum\limits_{n} a_{n}(-1)^n L_n^{{1\over
 2}}({{p^2\over{2\sigma^2}}})
\end{equation}
  The procedure is a smoothing operation, which works with noisy data.
   and which also allows for the inclusion of small corrections to the impulse
 approximation\cite{rmn}.  The errors in the measured $n(\vec p)$ are
 determined by the uncertainty in the the measured
  coefficients,through their correlation matrix, which is
 calculated by the fitting program. The coefficient $a_1$ is set to zero to avoid
 redundancy with the variation of $\sigma$. As a consequence, in units in which $\sigma$ is measured in
inverse Angstroms, and the energy is expressed in milli-electron
volts, the total kinetic energy(for a proton) is
K.E.=6.2705$\sigma^2$, even for strongly anharmonic momentum
distributions . When fit in the way described above, which we will
call a free fit,\cite{ds} since there is no model assumed, we find
that in fact the data can be fit well with  two coefficients that
are statistically significant, $a_2$ and $a_3$, and it is these
together with $\sigma$ that we present in the Table I below to
describe n(p).For more details on the fitting procedure see
Ref.\cite{rmn}.

The room temperature water and poly-crystalline ice data were
taken in standard aluminum sample holders, 10cm by 10cm by 1mm,
thin enough to lead to small multiple scattering which is
corrected for in all cases. A cell was designed specifically for
the high temperature  and high pressure measurements. The
background from this ZrTi cell appears as inelastic scattering and
is readily subtracted.\cite{rmn} It can provide pressure up to
2000 bar and temperature up to $450^oC$. The sample size in the
cell is 7 mm in diameter and 30 mm in height, presenting
considerably less sample to the beam than the water and ice
measurements, and hence leading to poorer statistics.  The heaters
and temperature sensors were all inserted in the cell. A 1 mm
steel pipe leads to an external water-pressurizer (i.e. a pump)
which provided the required pressure. The water used as pressure
transfer media and the sample volume was distilled H$_2$O.

\begin{table}
\caption{Parameters for Free Fit}
\begin{tabular}{||c|c|c|c||}
 Water Sample& $\sigma (\AA^{-1})$ &$a_4$& $a_6$ \\
\hline
Ice -4C&4.579&$.060\pm .0014$&$-.068\pm .016$\\
\hline
     Water 23C&4.841&$.185\pm .012$&$.015\pm.015$\\
\hline
T=400C, P=750bar&6.363&$.271\pm.022$&$.157\pm .028$\\
\hline
T=400C, P=750bar in C60  & 6.439 &$.592\pm.054 $& $.007\pm.07 $\\
\end{tabular}
\end{table}
This representation should be regarded as the data for n(p)
determined by the measurement. We show in Fig. 1 a comparison of
the free fits to the data for ice, water at room temperature,
supercritical water, and supercritical water contained in the
interstices of a C60 powder.  The quantity $4 \pi p^2 n(p)$, the
radial momentum distribution is presented, in order to compare
quantities with the same normalization. Although the temperature
has a profound effect on the structure, for a given structure, it
has only a small direct effect on the momentum distribution. The
corrections to the momentum widths for the ice due to  the
excitation of higher vibrational levels
 is negligible along the bond and only a few percent
transverse to the bond. What we are seeing is nearly entirely the
ground state momentum distribution, reflecting the average local
structure of the proton environment.

\begin{figure}\hspace{.1 \hsize}
\psfig{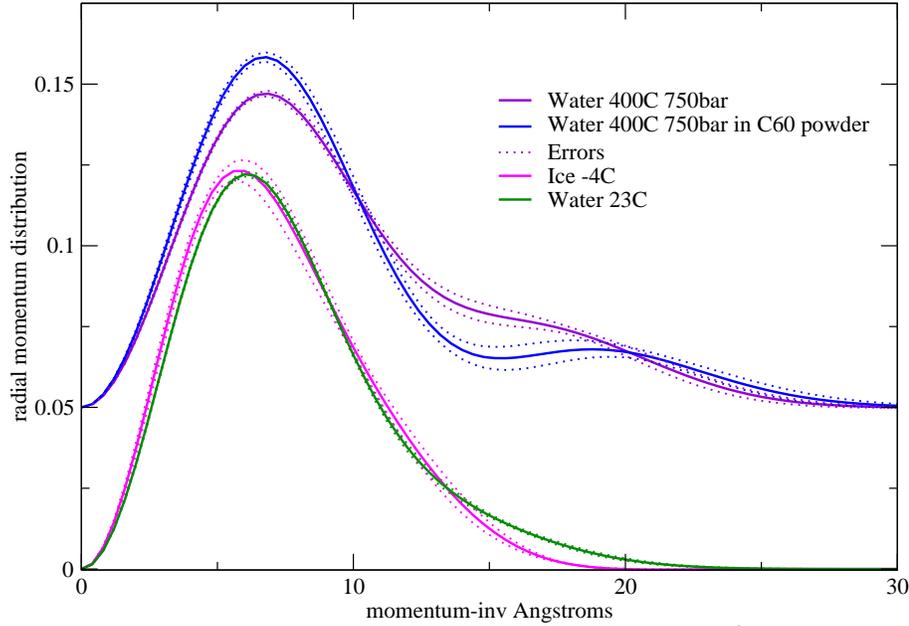}
 \caption{Comparison of fits to the data of the radial
momentum distribution, 4$\pi p^2 n(p)$  for a range of conditions
from ice to supercritical water in the interstices of a C60
poly-crystalline powder(typical size of pores 100 \AA)
corresponding to increasing disorder in the hydrogen bond network.
The supercritical water data has been displaced upward by .05 for
clarity. The density of the supercritical water is .65 gr/cc. }

\end{figure}
Structural  changes  in going from ice to water are noticeable in
the data. The tail of the distribution is due to the momentum
along the bond direction, since the proton is most tightly bound
in this direction, and tight binding implies high momentum width.
It is clear that this width
 has increased, which can be interpreted as due to the slight
increase in the average hydrogen bond distance in water, so that
the proton becomes more tightly bound to its covalently bonded
oxygen. The change in the covalent bond length is only 4$\%$, but
the measurements are clearly accurate enough to see this shift.
 This interpretation can also be applied to the
supercritical water data.
 The bonds are expected to be significantly longer or missing
altogether explaining the greatly increased momentum width.
However, the second peak in that data, is entirely unexpected from
a simple picture of the bond.  One could try to interpret the
second peak as a small population of protons with unusually high
momentum. However, this would mean an RMS energy of these protons
of approximately 2.5 ev, and a localization length of about .025
Angstroms. We know of no mechanism for producing such energetic
protons, and will not consider this possibility further. We will
interpret the data in terms of a single particle in an effective
potential due to its neighbors. With this interpretation, the
second peak  indicates that the proton is coherent over two
separated sites along the bond. This can be made clear by fitting
the data phenomenologically with a model, in which, in a frame of
reference where an individual bond is taken to lie along the z
axis, the motion transverse to the bond is harmonic and along the
bond given by a distribution that corresponds in real space to two
Gaussians separated by a distance d. We have then
\begin{equation}
n(p_x,p_y,p_z)={{2cos^2({{p_z d}\over{2\hbar}})}\over{ 1+e^{{-2d^2
\sigma_z^2} \over { \hbar^2}}}}
 \prod\limits_i {e^{-{p_i^2\over 2\sigma_i^2}}\over
(2\pi\sigma_i)^{1\over 2}} \label{dhp}
\end{equation}
 We will assume $\sigma_x=\sigma_y$. We have done fits with this condition relaxed,
 and find that it is well satisfied.  The parameter
$\sigma_z$ gives the width of the Gaussians in real space  through
the uncertainty relation. In the case that d=0, we have an
anisotropic Gaussian momentum distribution.

This distribution is then averaged over all angles, the
corresponding J(y) fit to the data , and the parameters of the
model determined.

We show in fig 2 the fits to that model, and in table 1 the
parameters obtained for those fits. The ice data is very
accurately described by an anisotropic Gaussian. The parameters of
the Gaussian correspond to vibrational energies in the transverse
and longitudinal directions of 105mev and 332mev respectively.The
water data is also well described this way, with a higher stretch
frequency of $367$ meV and no change in the transverse frequency,
although there are clearly additional anharmonicities that make
small corrections that are visible in the figure. These may be due
to a variation of the effective potential from site to site in the
water, or to an intrinsic anharmonicity in a single bond, perhaps
the precursor to the strong anharmonicity seen in the
supercritical water data. Indeed, if the distribution were due
entirely to a rotationally averaged gaussian, the coefficient of
$a_3$ would have to be negative, and it is not\cite{gr2}.  Li et
al\cite{li2} have measured the frequency of vibration for hydrogen
impurities in D$_2$O ice, which should be comparable to the
frequencies we infer from the anisotropic harmonic fit to our
data. They find that the two transverse vibrations of the proton
are at 105 meV and 200 meV, with the stretch mode at 405 meV.
The additional contributions to the anharmonic coefficients may be
responsible for the discrepancy with Li et al's results, since the
difference in the transverse mode frequencies we obtain by fitting
a Gaussian with three different vibrational energies is very
sensitive to the value of $a_3$.

The fits to the supercritical data require non zero values for the
parameter d giving the separation of minima in the potential
wells. That is the proton appears to be coherent over sites
separated by a distance of approximately .3 angstroms that are
both local minima of the potential.

\begin{table}
\caption{Parameters for Model Fit}
\begin{tabular}{||c|l|c|c||}

 Water Sample& $\sigma_z$(Inv. Angstroms) &$\sigma_x$(Inv. Angstroms)& d(Angstroms) \\
\hline
Ice -4C&$6.29\pm .51$&$3.53\pm .31$&0\\
\hline
     Water 23C&$6.73\pm .08$&$3.51\pm .04$&0\\
\hline
T=400C, P=750bar&$8.40\pm.19$&$5.70\pm.16$&$.316\pm.0045$\\
\hline
T=400C, P=750bar in C60  & $9.47\pm.26$ &$5.025\pm.11 $& $.274\pm.0043 $\\
\end{tabular}

\end{table}

\begin{figure}\hspace{.1 \hsize}
\psfig{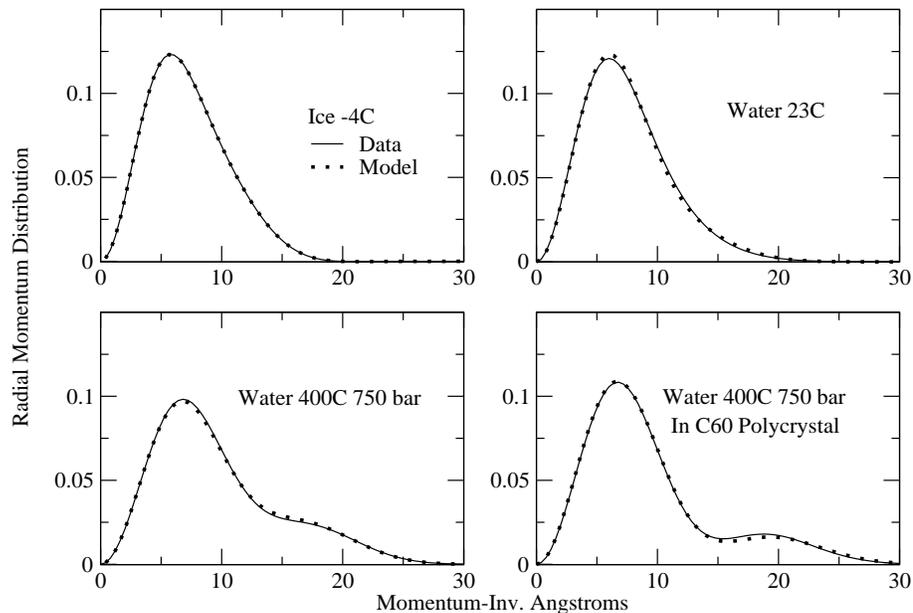}
 \caption{Comparison of radial
momentum distribution, 4$\pi p^2 n(p)$ with a one-particle model
based on double wells along the bond direction, Eq. \ref{dhp}  .
The additional peaks in the supercritical water data are due to
coherence over wells separated by distances of the order of .3
Angstroms in this model. There is only one well for the room
temperature water and the ice data.}

\end{figure}
We know of no prediction of such an effect. The disorder in the
hydrogen bond network would lead to bifurcated bonds,
 bent bonds double bonds  and missing bonds. It would seem that these would lead to tunnelling
motions transverse to the bond. In our experiment, however, the
tunnelling, or more precisely, the coherence, shows up along the
axis with a high momentum width, which is surely the axis of
stretching of the covalent bond.  To the extent that there are
linear bonds, this would be the bond axis. In fact, it is
reasonable to think that the supercritical water is made up of
small clusters that combine and break up on a time scale much
longer than our observation time, so we are seeing a snapshot
average of the ground states for the proton in these small
clusters.\cite{boero}  It is possible then, that what we are
seeing is the effect of cooperative tunnelling between single
bonds through the intermediate state of a bifurcated bond , as
observed in trimers and small clusters \cite{sayk}. Although the
cooperative tunneling motion in these small clusters involves
primarily the transverse motion, this could be accompanied, due to
a coupling of the longitudinal and transverse modes,  by changes
in the momentum distribution along the bond direction, which is
what we see. We note in this regard that the interpretation in
terms of a single particle effective potential, although it fits
well, is only a phenomenological representation of a more complex
many-body phenomenon. We also note that a double well potential
has been posited as an explanation for the extroardinarily large
Debye-Waller factors observed in water in the interior of carbon
nanotubes,\cite{sacha} suggesting that it is the disorder of the
network that is the origin of the double well potential there too.

The wells are sufficiently separated that the wave-function
actually becomes bimodal. We show in fig 3 the probability, (the
wave-function squared) corresponding to the fitted momentum
distributions for three of the  measurements, together with the
potential that would produce that wavefunction.
\begin{figure}\hspace{.1 \hsize}
\psfig{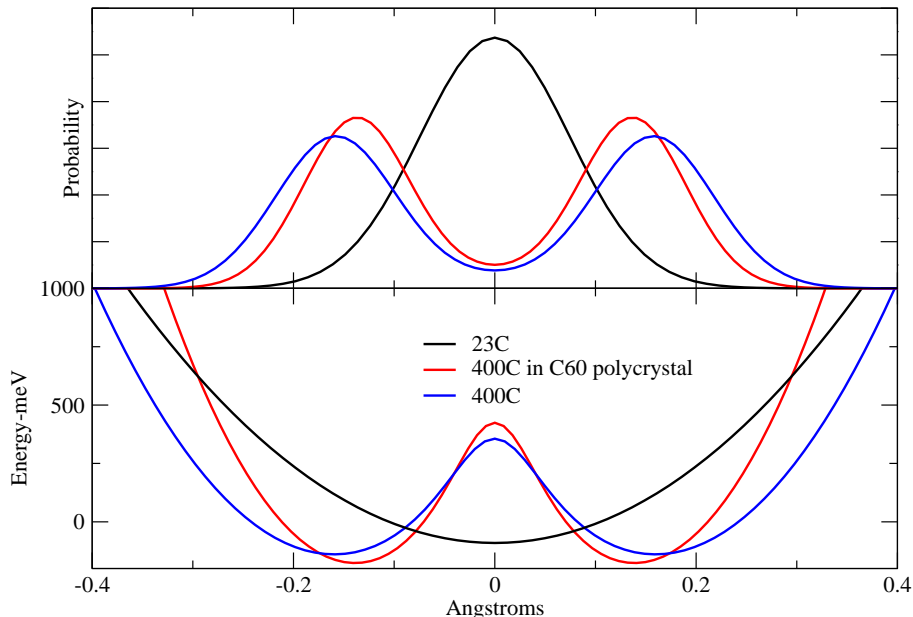}
 \caption{Comparison of probability,
 based on an effective single particle model, for finding a proton at a position along the bond.
  The zero position is unknown from our experiments, so that the coordinate is relative
  to the most probable position.
   The ice and room temperature results are Gaussian, corresponding to harmonic wells.
   The high temperature wavefunctions are assumed to be the sum of two Gaussians separated
    by some distance(Table II), and the potential is that which would produce those wavefunctions. }

\end{figure}

 One can
expect that the the hydrogen bond network will be even more
distorted for the water in the interstices of the powdered C60
than in the pure water, and we will be seeing an average over
bonds that are near the surface with those that are relatively "in
the bulk". The size of the surface layer at ordinary temperatures
is in some dispute\cite{n1,n2}, ranging from 1.5-5nm and it seems
certain that it would be significantly larger for the
supercritical water. It is conceivable, given the size of the
interstices in the C60 poly-crystal, that the entire volume of
water would be strongly affected by the contact with the surfaces.
Consistent with this, we find that the protons are more localized
in the C60 interstices than in the "free" supercritical water.

In conclusion, the results shown here, in addition to providing a
detailed picture of the dynamics of the proton in water, show that
the momentum distribution of the proton  can be measured with
sufficient accuracy to provide detailed information about the
local structure, even in liquids or powder samples.

\begin{acknowledgements} We would like to thank Frank Stillinger, Tony Haymet and Ariel Chialvo for
useful discussions. This work was supported by DOE Grant
DE-FG02-03ER46078
\end{acknowledgements}\\\\

\vfill\eject


\begin{references}
\bibitem{gr2}The average may be done analytically. For
the case that the two transverse directions have the same
variance, we find $<J(y)>$ is of the form given in Eq.\ref{exp},
with $a_1=0$, $\sigma^2=(2\sigma_x^2+\sigma_z^2)/3$, and
$a_n$=$({\delta\sigma^2\over \sigma^2})^n<(1/3-cos^2(\theta))^n>$
, where $\delta\sigma^2=(\sigma_x^2- \sigma_z^2)$.   Explicitly,
$a_2$=$4/45({\delta\sigma^2\over \sigma^2})^2$ and
$a_3=16/945({\delta\sigma^2\over \sigma^2})^3$
\end{references}
\end{document}